\documentclass[a4paper,11pt]{article}
\usepackage{pos}

\title{$\tau_{B_c}$ in the Standard Model}

\author*[a]{Jason~Aebischer}
\author[b]{Benjam\'in Grinstein}

\affiliation[a]{Physik-Institut, Universit\"at Z\"urich, CH-8057 Z\"urich, Switzerland}

\affiliation[b]{Department of Physics, University of California at San Diego,
    La Jolla, CA 92093, USA}

\emailAdd{jason.aebischer@physik.uzh.ch}
\emailAdd{bgrinstein@ucsd.edu}

\abstract{The Standard Model prediction of the $B_c$ lifetime is discussed, together with the dominant uncertainties and strategies on how to improve them. Furthermore, a new method to compute the $B_c$ lifetime based on the operator product expansion is proposed. It relies on differences of $B,\,D$ and $B_c$ meson decay rates, in which the free-quark contributions cancel out, reducing the uncertainty of the theory prediction.}

\FullConference{%
  Corfu Summer Institute 2021 "School and Workshops on Elementary Particle Physics and Gravity"\\
  29 August - 9 October 2021\\
  Corfu, Greece
}

\begin{document}
\maketitle

\section{Introduction}

\noindent
The $B_c=(\overline b c)$ meson is an interesting particle to study, since it contains two different heavy quarks. This makes it a prime candidate for the use of Non-Relativistic QCD (NRQCD), in which an expansion of the heavy quark velocities $v_b$ and $v_c$ can be carried out. Together with the operator production expansion (OPE) approach this has lead to the most precise theory prediction of the $B_c$ lifetime \cite{Beneke:1996xe,Bigi:1995fs,Chang:2000ac}. Other approaches include QCD Sum Rules \cite{Kiselev:2000pp} as well as Potential models \cite{Gershtein:1994jw}, which lead to comparable results.

\noindent
Another interesting fact about the $B_c$ meson is that its lifetime puts stringent constraints on New Physics models containing new scalars, that are able to explain the current $R(D)$ and $R(D^*)$ anomalies. Prominent examples are scalar Leptoquarks and Two-Higgs-Doublet models \cite{Alonso:2016oyd,Blanke:2018yud,Blanke:2019qrx}.

\noindent
From the experimental point of view the lifetime of the $B_c$ is very precisely measured to be

\begin{equation}\label{eq:exp}
  \tau_{B_c}^{\text{exp}} = 0.510(9)\text{ps}\,,
\end{equation}

\noindent
which is the averaged value of the LHCb \cite{LHCb:2014ilr,LHCb:2014glo} and CMS \cite{CMS:2017ygm} measurements.

\noindent
This precision is however not matched from the theory side, due to several large uncertainties. They result for instance from neglecting higher order non-perturbative corrections, parametric uncertainties as well as the inclusion of the strange quark mass. The main uncertainties stem however from the treatment of the masses of the quarks inside the $B_c$. To examine this behaviour in more detail, three different mass schemes were studied in \cite{Aebischer:2021ilm,Aebischer:2021eqy} to compute the $B_c$ decay rate in the OPE approach, which will be discussed below together with the other uncertainties.

\noindent
The rest of the article is organized as follows: In sec.~\ref{sec:mass} the three mass schemes are presented, together with the corresponding predictions of the $B_c$ decay rate. In sec.~\ref{sec:uncertainties} we discuss the uncertainties occurring in the OPE approach. A new method to compute the $B_c$ decay rate is introduced in sec.~\ref{sec:newmethod}, before we summarize in sec.~\ref{sec:summary}.

\section{Mass schemes}\label{sec:mass}

\noindent
In this section we summarize the three different mass schemes, which were used in \cite{Aebischer:2021ilm} to determine the $B_c$ decay rate. They include the $\overline{\text{MS}}$, the Upsilon and the meson scheme.

\subsection{$\overline{\text{MS}}$ scheme}

\noindent
In the $\overline{\text{MS}}$ mass-scheme the on-shell (OS) masses of the $\overline{b}$ and $c$ quarks are expressed in terms of the renormalized $\overline{\text{MS}}$ masses via the following equation:

\begin{equation}
  \label{eq:poleM1loop}
  m_q=\overline{m}_q(\mu)\left[1+\frac{\alpha_s(\mu)}{\pi}\left(\frac{4}{3}-\ln\left(\frac{\overline{m}_q(\mu)^2}{\mu^2}\right)\right)\right]+\mathcal{O}(\alpha_s^2)\,.
\end{equation}

\noindent
In our computation we use the lattice results \cite{Bazavov:2018omf,Colquhoun:2014ica,Lytle:2018evc} for the $\overline{\text{MS}}$ masses, which lead to the following decay rate of the $B_c$:

\begin{equation}\label{eq:MSbar}
  \Gamma^{\overline{\text{MS}}}_{B_c} = (1.51\pm 0.38|^{\mu}\pm 0.08|^{\text{n.p.}}\pm 0.02|^{\overline{m}}  \pm0.01|^{m_s}\pm 0.01|^{V_{cb}})\,\,\text{ps}^{-1}\,,
\end{equation}
where the third uncertainty is due to the $\overline{\text{MS}}$ masses. The other uncertainties will be discussed in the following section. The value obtained in \eqref{eq:MSbar} is to be compared with the experimental value of the decay rate, given by

\begin{equation}\label{eq:Gexp}
  \Gamma_{B_c}^\text{exp} = 1.961(35) \,\text{ps}^{-1}\,.
\end{equation}

\subsection{Upsilon scheme}

\noindent
In this mass scheme, the OS mass of the $\overline b$ quark is expressed in terms of the very precisely measured Upsilon 1S state, by using the relation \cite{Pineda:1997hz,Melnikov:1998ug}

\begin{equation}
  \frac{\tfrac12m_{\Upsilon}}{m_b}=1-\frac{(\alpha_s C_F)^2}{8}
  \left\{1
    +\frac{\alpha_s}{\pi}\left[\left(\ln\left(\frac{\mu}{\alpha_sC_Fm_b}\right)+\frac{11}{6}\right)\beta_0-4\right]^2+\cdots\right\}\,,
\end{equation}
where $\beta_0$ is the one-loop beta function factor of the strong coupling constant. A similar relation is used to express the charm quark mass in terms of the $J/\Psi$ mass. In our analysis we use the PDG values $m_{\Upsilon}=9460.30(26)$ MeV and $m_{J/\Psi}=3096.900(6)$ MeV \cite{Tanabashi:2018oca}, which gives a $B_c$ decay rate of

\begin{equation}\label{eq:upsG}
  \Gamma^{\text{Upsilon}}_{B_c} = (2.40\pm 0.19|^{\mu}\pm 0.21|^{\text{n.p.}} \pm0.01|^{m_s}\pm 0.01|^{V_{cb}})\,\,\text{ps}^{-1} \,,
\end{equation}
where the uncertainties of $m_{\Upsilon}$ and $m_{J/\Psi}$ are completely negligible.

\subsection{Meson scheme}

\noindent
As a third scheme we use the so-called meson scheme, where the OS quark masses are expressed in terms of the meson masses by use of the HQET relation

\begin{equation}
  \label{eq:poleMassDiff}
  m_b-m_c=\overline{m}_B-\overline{m}_D+\frac12\lambda_1\left(\frac1{m_b}-\frac1{m_c}\right)
  +\cdots
\end{equation}

\noindent
where $\lambda_1=-0.27 \pm 0.14$ \cite{Hoang:1998ng}, and $\overline{m}_B=\frac14(3 m_{B^*}+m_B)$ and $\overline{m}_D=\frac14(3 m_{D^*}+m_D)$ denote the spin and isospin-averaged meson masses. In this scheme we obtain

\begin{equation}\label{eq:mesonG}
  \Gamma^{\text{meson}}_{B_c} = (1.70\pm 0.24|^{\mu}\pm 0.20|^{\text{n.p.}} \pm0.01|^{m_s}\pm 0.01|^{V_{cb}})\,\,\text{ps}^{-1} \,,
\end{equation}
where the obtained value is in rather good agreement with the measurement in eq.~\eqref{eq:Gexp}.

\section{Uncertainties}\label{sec:uncertainties}

\noindent
In this section we discuss the uncertainties of the theoretical prediction. A more detailed analysis can be found in \cite{Aebischer:2021ilm}.

\subsection{Scale dependence}

\begin{figure}
  \begin{center}
    \includegraphics[width=0.45\textwidth]{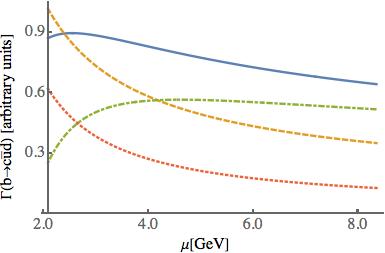}\hspace{0.05\textwidth}
    \includegraphics[width=0.45\textwidth]{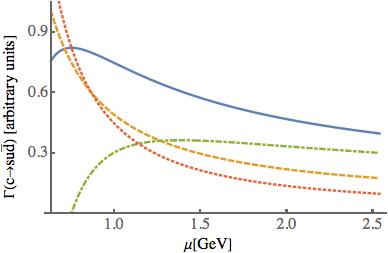}
    \caption{\label{fig:mu-dep} Scale dependence of the LO decay rates $\Gamma(b\to cud)$ (left panel) and
      $\Gamma(c\to sud)$ (right panel) in the $\overline{\text{MS}}$ scheme. The NLO (solid-blue) and LO (dashed-orange) calculations are shown, respectively. The LO calculation to which the term with
      the explicit factor of $\alpha_s\ln(\mu)$ in the NLO decay rate is added is shown in green, displaying cancellation of
      scale dependence at $\mathcal{O}(\alpha_s)$. The NLO decay rate omitting the term with an explicit factor of
      $\ln(\mu)$ is given by the dotted-red line.}
    \end{center}
\end{figure}

\noindent
The residual renormalization-scale dependence from truncating the loop expansion is the largest uncertainty in the $B_c$ lifetime. It enters mainly through the OS mass replacements of the quarks in the three different schemes, since these relations are only used at the one-loop level. The scale dependence is largest in the $\overline{\text{MS}}$ scheme, which is illustrated in Fig.~\ref{fig:mu-dep}: It depicts the scale dependence of the leading order (LO) quark decay rates $\Gamma(b\to cud)$ and $\Gamma(c\to sud)$.

\noindent
To reduce the scale dependence in our results, higher order QCD corrections have to be incorporated in the calculation, first in the OS mass relations and secondly also in free-quark decay rates.

\subsection{Non-perturbative uncertainties}

\noindent
Further uncertainties result from the NRQCD expansion in the quark velocities $v_b$ and $v_c$, which has been truncated at $\mathcal{O}(v^4)$. Furthermore, the non-perturbative (n.p.) parameters also have uncertainties which are incorporated in the n.p. uncertainty estimations in eqs.~\eqref{eq:MSbar}, \eqref{eq:upsG} and \eqref{eq:mesonG}. The main improvement in these uncertainties would be to include higher-order corrections in the velocity expansion. It would however also be favourable to have lattice results available for the n.p. parameters.

\subsection{Parametric uncertainties}

\noindent
Additional uncertainties result from all the parameters that are involved in the calculation, the largest one stemming from the uncertainty of the CKM matrix element $V_{cb}$ given in the last uncertainties of eqs.~\eqref{eq:MSbar}, \eqref{eq:upsG} and \eqref{eq:mesonG}. In the $\overline{\text{MS}}$ scheme also the $\overline{\text{MS}}$-masses introduce a rather large uncertainty, which is shown in third uncertainty in eq.~\eqref{eq:MSbar}.

\subsection{Strange quark mass}

\noindent
 In the spectator $c$-decays a non-vanishing strange quark mass reduces the decay rate by about 7\% in the three different mass schemes. The introduced uncertainty when neglecting $m_s$ in the $\bar b$-quark decay can be estimated naively by considering the factor $(m_c/m_b)^2\sim0.1$, multiplied by the corresponding decay rate and a factor of 7\%. In the $c$-quark decays the parametric uncertainty resulting from $\overline{m}_s(2\;\text{GeV})$ leads to an uncertainty of $\Delta\Gamma_c\sim0.01\;\text{ps}^{-1}$.

\section{Novel determination of $\Gamma_{B_c}$}\label{sec:newmethod}

\noindent
To reduce the rather large uncertainties in the theory prediction, which mainly result from the scale dependence, we will adopt a novel method to compute the $B_c$ decay rate, first described in \cite{Aebischer:2021eio}. The idea is to make use of the non-perturbative expansion of the decay rate not only for the $B_c$ meson, but also for the $B$ and $D$ mesons, by considering the combination

\begin{align}\label{eq:diff}
  \Gamma(B)+\Gamma(D)-\Gamma(B_c) &= \Gamma^{n.p.}(B)+\Gamma^{n.p.}(D)-\Gamma^{n.p.}(B_c) \nonumber \\
  &+\,\Gamma^{\text{WA}+\text{PI}}(B)+\Gamma^{\text{WA}+\text{PI}}(D)-\Gamma^{\text{WA}+\text{PI}}(B_c)\,.
\end{align}

\noindent
where the rates on the left-hand side are given by

\begin{equation}\label{eq:GM}
  \Gamma(H_Q) = \Gamma_Q^{(0)}+\Gamma^{n.p.}(H_Q)+\Gamma^{\text{WA}+\text{PI}}(H_Q)+\mathcal{O}(\frac{1}{m_Q^4})\,,
\end{equation}

\noindent
for a meson $H_Q$ with heavy quark $Q$ and where WA and PI stand for Weak Annihilation and Pauli Interference contributions. On the right-hand side of eq.~\eqref{eq:diff} the LO quark decay rates $\Gamma^{(0)}_Q$ drop out, since they are independent of meson states. Therefore the largest source of scale dependence vanishes, which reduces the error of the result. In order to determine the $B_c$ decay rate eq.~\eqref{eq:diff} can be applied for either charged or neutral $B$ and $D$ mesons, resulting in four different ways to compute $\Gamma(B_c)$. The results using these four different combinations are given in Tab.~\ref{tab:res}.

\begin{table}[t]
\centering
 \begin{tabular}{|l |c |c |c |c|}
 \hline
 & $B^0,D^0$ & $B^+,D^0$ & $B^0,D^+$ & $B^+,D^+$ \\ [0.5ex]
 \hline \hline
$\Gamma^{\text{meson}}_{B_c}$ & 3.03 $\pm$ 0.51 & 3.03 $\pm$ 0.53 & 3.33 $\pm$ 1.29 & 3.33 $\pm$ 1.32 \\
 \hline
 \end{tabular}
 \caption{\small
Results obtained for $\Gamma^{\text{meson}}_{B_c}$ in units of $\text{ps}^{-1}$ in the meson scheme, using different combinations of $B$ and $D$ mesons in eq.~\eqref{eq:diff}.
}
  \label{tab:res}
\end{table}

\noindent
Several reasons for the disparity between the obtained results and the experimental value in eq.~\eqref{eq:Gexp} can be brought forward:

\begin{itemize}
  \item Underestimation of the uncertainties from NLO corrections
  \item Eye-graph contributions to matrix elements, generally neglected in lattice computations \cite{Becirevic:2001fy}
  \item Effects of dimension-seven contributions to charm decays \cite{King:2021xqp}
  \item Quark-hadron duality violation
\end{itemize}

\noindent
It is certainly worth investigating all the above mentioned points in more detail. In case the first three options fail to explain the discrepancy, the last one has to be seriously considered when describing meson decay rates.

\section{Summary}\label{sec:summary}

\noindent
We have presented an updated analysis of the $B_c$ decay rate, following the OPE approach. Three different mass schemes have been studied, which all lead to results in agreement with experiment and with each other. Furthermore an analysis of the theory uncertainties has been presented, where the scale-dependence makes up most of the total uncertainty.

\noindent
In a second part a new method to determine the $B_c$ lifetime has been presented, which relies on taking differences of $B,\,D$ and $B_c$ decay rates. The rather large values obtained in this method might have several reasons, including underestimation of uncertainties, unknown eye-graph contributions, dimension-seven contributions to the charm decays or even quark-hadron duality violation.

\small
\bibliographystyle{JHEP}
\bibliography{AGbib}

\end{document}